\documentclass[journal]{IEEEtran}
\usepackage{braket}
\usepackage{ amsmath, amssymb, amsfonts} 
\usepackage{amsthm}
\usepackage{CJK}
\usepackage{indentfirst}
\usepackage{cases}
\usepackage{cite}
\usepackage{epsf, epsfig}
\usepackage{stfloats}
\usepackage{mathtools}
\usepackage{siunitx}
\usepackage{enumitem}
\usepackage{bm}
\usepackage{amsfonts,amsmath}
\usepackage{graphicx}
\usepackage{subfigure}
\usepackage{multirow}
\usepackage{color}
\usepackage{amssymb}
\usepackage{algorithm}
\usepackage{setspace}
\usepackage{cases}
\usepackage{cite}
\usepackage{cuted}
\usepackage{booktabs}
\usepackage{caption}
\usepackage{epsf, epsfig}
\usepackage{amsfonts}
\usepackage{graphicx}
\usepackage{subfigure}
\usepackage{multirow}
\usepackage{color}
\usepackage{amssymb}
\usepackage{setspace}
\usepackage{gensymb}
\usepackage{bbm}
\usepackage{makecell}
\usepackage{textcomp}
\usepackage{url}
\usepackage{algorithm}
\usepackage{graphicx}
\usepackage{subcaption}
\usepackage{algpseudocode} 
\usepackage[colorlinks,citecolor=red, linkcolor=blue, anchorcolor=green]{hyperref}

\allowdisplaybreaks

\begin{document}
\title{Geometry-Aware Networking for Low-Altitude Economy:
Movable Antennas in Space–Air–Ground Integrated Systems}

\author{
	
\IEEEauthorblockN{ 
Heyou Liu,~\IEEEmembership{Graduate Student Member,~IEEE}, 
Bang Huang,~\IEEEmembership{Member,~IEEE},
Mohamed-Slim Alouini,~\IEEEmembership{Fellow,~IEEE}
\thanks{
        \IEEEauthorblockA{Heyou Liu, Bang Huang and Mohamed-Slim Alouini are with Computer, Electrical, and Mathematical Science and Engineering Division, King Abdullah University of Science and Technology, Thuwal, Saudi Arabia.} \newline
         (e-mail: \{heyou.liu, bang.huang,  slim.alouini\}@kaust.edu.sa.) (Corresponding author: Bang Huang)}
}
}

\maketitle
\begin{abstract}
Space--air--ground integrated networks (SAGINs) are emerging as a key foundation for future non-terrestrial networks (NTNs) and low-altitude economy services. 
However, their performance is increasingly limited not only by communication resources, but by the inability to adapt to rapidly changing spatial geometry. Here, spatial geometry refers to the relative configuration among network nodes, obstacles, and targets, which directly determines propagation conditions, blockage states, interference patterns, and sensing observability.
This trend becomes more pronounced as low-altitude operations grow in density and complexity, causing the dominant bottleneck to shift from static resource allocation toward real-time maintenance of favorable spatial geometry across layers.
In this article, we argue that movable antenna (MA) technology provides a fundamentally new perspective for SAGIN design. By enabling controlled antenna displacement, MA introduces a spatial degree of freedom that allows the network to directly adapt local spatial geometry at fine granularity, rather than passively reacting to it through beamforming or platform mobility.
We present a geometry-aware, layered SAGIN architecture, where Low-Earth-Orbit (LEO) provides macro-scale coverage and coordination, High-Altitude Platform Stations (HAPS) enables regional continuity and backhaul support, and MA is incorporated into the layered design to enable fine-grained geometry adaptation, particularly at unmanned aerial vehicles (UAVs) and terrestrial layers where local channel dynamics are most pronounced. We further discuss how such geometry control enhances robustness, supports multi-functional operation spanning communication, sensing, control, and navigation, and enables more flexible spatial cooperation across layers.
Overall, this work positions MA-enabled SAGIN not merely as an antenna enhancement, but as a paradigm shift toward geometry-aware, multi-scale, and multi-functional network design for future 6G and beyond systems.
\end{abstract}


\section{Introduction}

The rapid emergence of low-altitude economy (LAE) services, such as aerial logistics, urban air mobility, infrastructure inspection, and emergency response, is fundamentally reshaping the requirements of future wireless networks. On the one hand, LAE unlocks unprecedented opportunities by enabling flexible, on-demand, and large-scale aerial services. On the other hand, it introduces a set of {new and stringent system challenges}, including highly dynamic network topology, severe and intermittent blockage, and tightly coupled communication, sensing, control, and navigation requirements under strict latency and reliability constraints \cite{zeng2019accessing,mahboob2024revolutionizing}.
Unlike conventional terrestrial systems, LAE applications operate in complex three-dimensional environments, where the {spatial geometry of the network evolves rapidly} due to UAV mobility, platform instability, and urban obstructions.
In this article, spatial geometry refers to the relative positions, orientations, and blockage relationships among communication nodes, obstacles, and sensing targets, which jointly determine propagation quality, interference exposure, and observability.
In such settings, system performance is no longer limited solely by communication resources such as bandwidth or power, but increasingly by the ability to maintain favorable {spatial conditions} for reliable signal propagation, accurate sensing, and safe control. This shift highlights a fundamental challenge: {how to adapt to fast and fine-grained geometry variations in real time \cite{liu2018space}}.

Space--air--ground integrated networks have been widely recognized as a promising architectural solution to support LAE services. By integrating low earth orbit (LEO) satellites, high-altitude platforms (HAPS), unmanned aerial vehicles (UAVs), and terrestrial infrastructure, SAGINs enable multi-scale coverage, improved resilience, and flexible service provisioning across large geographic regions \cite{xu2023space}. However, this architectural flexibility also amplifies the geometry challenge. The coexistence of multiple layers leads to highly dynamic spatial configurations, severe low-altitude blockage, and complex cross-layer interference patterns. As a result, a critical mismatch emerges between macro-scale network coordination and micro-scale geometry dynamics, which becomes a dominant performance bottleneck \cite{liu2018space,mahboob2024revolutionizing}.
Existing approaches attempt to address these challenges through beamforming, UAV trajectory optimization, or reconfigurable intelligent surfaces (RIS) \cite{qin2023coverage,wu2019towards, di2020smart}. While effective at certain scales, these techniques remain fundamentally limited in addressing {real-time, fine-grained geometry adaptation}. Beamforming operates on fixed antenna locations and cannot overcome unfavorable spatial configurations. UAV trajectory optimization provides large-scale flexibility but is constrained by mobility dynamics and slow adaptation timescales. RIS relies on environmental reconfiguration and is inherently limited by deployment constraints and lack of instantaneous adaptability. Consequently, none of these approaches can fully resolve the geometry mismatch inherent in LAE-enabled SAGINs.

Movable antenna technology offers a fundamentally different solution to this problem. By enabling controlled displacement of antenna elements within a local region, MA introduces a new spatial degree of freedom that allows the system to directly {manipulate its interaction with the surrounding geometry \cite{zhu2023modeling,zhu2025tutorial}}. This capability enables real-time adaptation to local spatial conditions, including blockage avoidance, interference reshaping, and geometry optimization for both communication and sensing tasks. 
More importantly, MA is uniquely positioned to address the core challenges of LAE. The very factors that make LAE difficult, such as rapid geometry variation, 3D interference, and strict reliability requirements, are precisely those that MA can effectively mitigate through fine-grained spatial control. This makes MA not merely a performance-enhancing technique, but a {critical enabler} for next-generation SAGINs \cite{zhu2025tutorial,shao2025tutorial}.

\begin{figure*}[htp]
\centering
\includegraphics[width=0.85\textwidth]{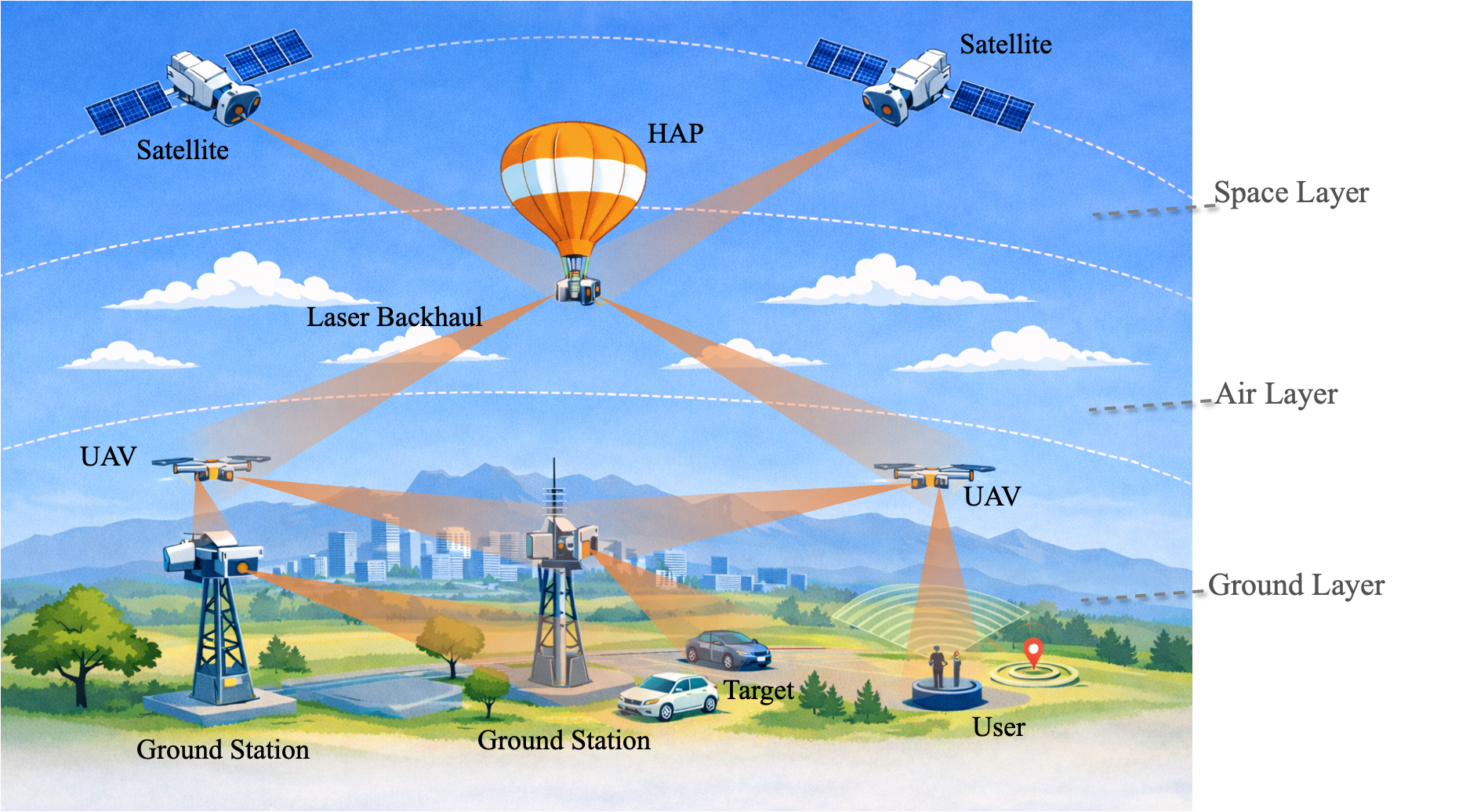}
\caption{
Illustration of space--air--ground integrated networks for low-altitude economy services. LEO satellites provide macro-scale coverage and global coordination, HAPS platforms enable regional continuity and backhaul support, while UAVs and terrestrial nodes operate in highly dynamic local environments requiring reliable communication, sensing, and navigation.
}
\label{fig:system_model}
\end{figure*}

\begin{figure}[htp]
    \centering
    \includegraphics[width=1\linewidth]{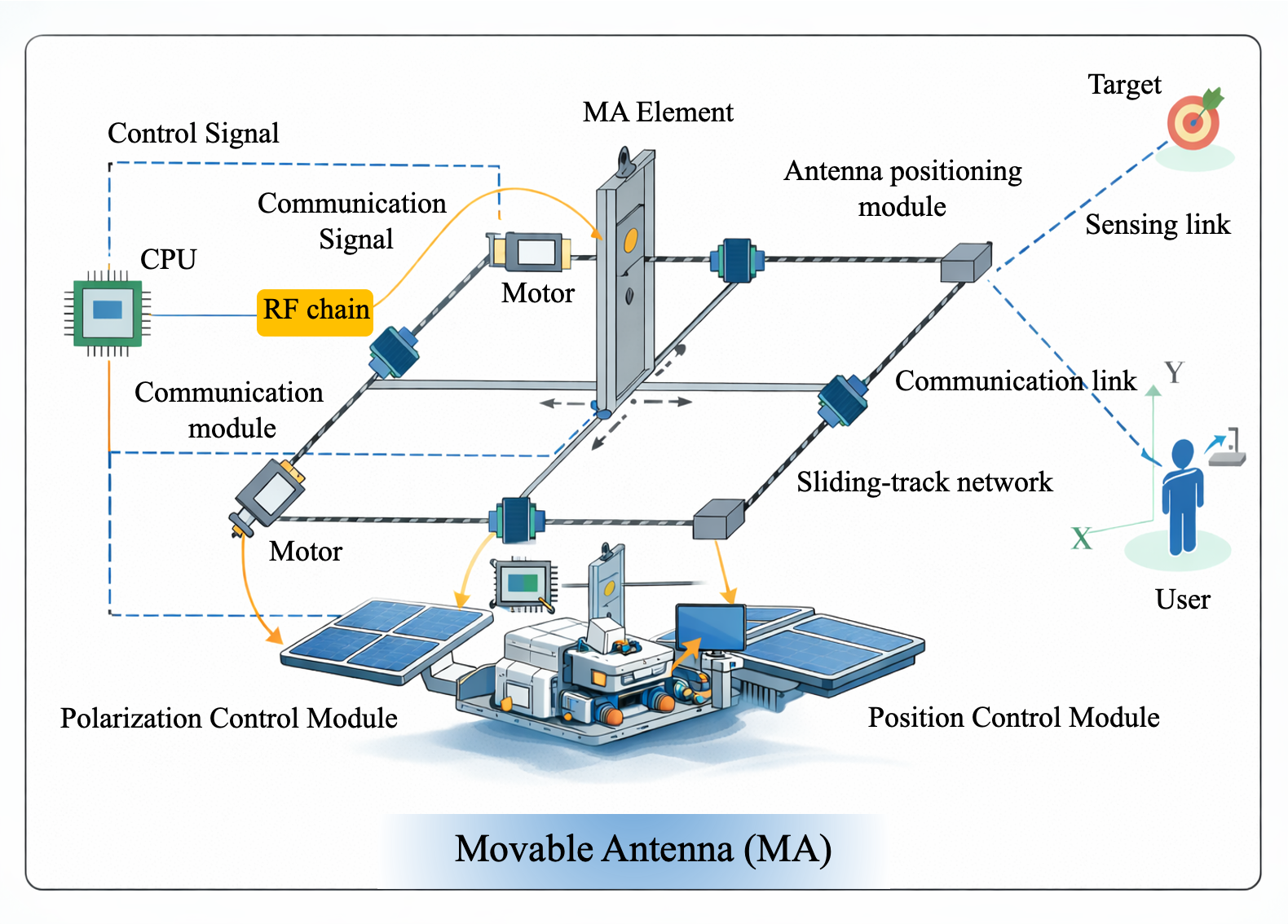}
    \caption{
MA architecture enabling joint communication and sensing. By dynamically adjusting antenna positions, MA provides fine-grained geometry control that can be integrated into SAGIN to mitigate blockage and improve link reliability in LAE scenarios.
}
    \label{fig:MA}
\end{figure}

In this article, we explore how MA technology can be systematically integrated into SAGINs to support LAE services. We discuss that MA should not be viewed merely as a link-level enhancement, but as a cross-layer mechanism that bridges macro-scale coverage control and micro-scale geometry adaptation.
To illustrate this vision, Fig.~\ref{fig:system_model} presents a representative SAGIN architecture for LAE scenarios, highlighting the multi-layer structure and the dynamic operating environment. However, as shown in Fig.~\ref{fig:system_model}, the performance of such systems is fundamentally constrained by the inability to adapt to rapidly changing spatial geometry. To address this limitation, Fig.~\ref{fig:MA} introduces the architecture of an MA system, which enables fine-grained, real-time geometry control through antenna mobility.
Building upon these insights, we first discuss why MA is essential in SAGINs, then present a layered architectural vision, analyze its impact on communication, sensing, control, and navigation, and finally highlight key challenges and future research directions. Our goal is to provide a unified perspective on how geometry-aware antenna technologies can fundamentally reshape future non-terrestrial networks. This shift suggests that future SAGIN design should move beyond communication-centric optimization toward geometry-aware system design, where spatial configuration becomes a first-class resource.

\section{Why Movable Antennas in SAGIN? From Connectivity-Centric to Geometry-Aware Networking}

\subsection{Limitations of Existing Spatial Adaptation Mechanisms}

The fundamental challenge in SAGINs does not merely lie in dynamic environments, but in the {mismatch between multi-scale network control and fast local geometry variation}, which becomes particularly severe in low-altitude scenarios.
Specifically, low-altitude networks exhibit several {unique and tightly coupled impairments}:

\begin{itemize}
    \item \textbf{Intermittent LoS blockage}: Although air-to-ground links are often LoS-dominant, they are highly vulnerable to sudden blockage caused by buildings, terrain, and UAV attitude variation, leading to abrupt link degradation.
    
    \item \textbf{Micro-scale geometry fluctuation}: UAV jitter, platform vibration, and rapid orientation changes introduce wavelength-scale variations in the effective channel, resulting in fast beam misalignment and unstable link quality.
    
    \item \textbf{3D interference exposure}: Unlike terrestrial systems, aerial nodes are simultaneously exposed to interference from multiple directions, making interference highly dynamic and difficult to suppress using conventional spatial filtering.
    
    \item \textbf{Coupled communication--sensing degradation}: In ISAC-enabled low-altitude systems, unfavorable geometry not only reduces communication quality but also deteriorates sensing accuracy due to poor observation angles and weak echo paths.
\end{itemize}

A natural question then arises as to whether existing techniques can effectively address these geometry-related challenges in SAGIN.
Existing approaches are fundamentally limited in their ability to provide real-time and fine-grained spatial adaptation. Beamforming enhances signal transmission but operates over fixed antenna structures and cannot alter unfavorable spatial configurations. UAV trajectory optimization provides large-scale geometry adjustment, yet it evolves at much slower timescales and is constrained by mobility, energy consumption, and mission requirements, making it incapable of responding to fast local fluctuations. RIS-based solutions reshape the propagation environment, but their effectiveness depends heavily on deployment conditions and they lack instantaneous adaptability to dynamic geometry changes.

As a result, current approaches fail to address a critical missing capability in current SAGIN designs, namely {real-time, fine-grained control over local spatial geometry}. This limitation becomes the dominant performance bottleneck in LAE services, where reliability, perception, and control must all be maintained under rapidly changing conditions. These limitations are not incidental, but stem from a fundamental inability to control spatial geometry at the link level, including link distance, blockage state, angular relationship, and observation condition.

\subsection{MA as a Geometry-Control Mechanism}

MA fundamentally differs from existing technologies by introducing direct and fine-grained control over the spatial geometry of wireless interaction. This capability enables a new form of adaptation that operates at a much finer scale than beamforming or platform mobility. Unlike beamforming, which adjusts signal weights over fixed antenna positions, MA allows the system to physically explore the local channel field. Unlike UAV trajectory control, which is constrained by mobility dynamics, MA operates at a much faster timescale and can respond to instantaneous geometry variations. Unlike RIS, which passively reshapes the environment, MA actively reconfigures the transceiver side. Here, MA refers to small-scale antenna displacement and orientation adjustment within a compact local region around the transceiver, rather than movement of the entire UAV or platform. The corresponding actuation range is typically local and aperture-limited, e.g., on the order of several to tens of wavelengths, which is sufficient to induce meaningful changes in blockage exposure, angular response, interference coupling, and sensing geometry without requiring platform-level repositioning.
This distinction is critical in low-altitude scenarios, since platform mobility and antenna mobility operate at fundamentally different spatial and temporal scales. By exploiting small-scale antenna displacement or orientation adjustment, MA can: i). compensate for beam misalignment caused by UAV jitter and platform instability, ii). rapidly recover from blockage by identifying favorable local propagation paths, iii). suppress 3D interference through spatial decorrelation and directional filtering, and iv). actively optimize sensing geometry for improved target observability and estimation accuracy.
From a system-level perspective, MA introduces a new capability that is absent in existing SAGIN designs, namely {real-time geometry control at the link level}. This capability directly targets the dominant impairments in low-altitude networks and cannot be achieved by conventional techniques alone.

\subsection{Bridging Multi-Scale Adaptation in SAGIN}

The significance of MA becomes more evident when viewed from a multi-scale perspective. SAGIN inherently operates across different spatial and temporal scales:

\begin{itemize}
    \item LEO provides macro-scale coverage and global coordination;
    \item HAPS offers regional persistence and service continuity;
    \item UAVs and terrestrial nodes handle localized access and rapid adaptation.
\end{itemize}

However, a fundamental gap exists between these layers. Upper layers can efficiently manage large-scale coverage and connectivity, but lack the ability to respond to fast local geometry variations. Lower layers, on the other hand, operate in highly dynamic environments but are constrained by limited spatial adaptation capabilities.

MA bridges this gap by introducing {micro-scale geometry adaptability} that complements macro- and meso-scale control. Specifically, MA enables fast, fine-grained spatial adjustment at the link level, allowing the system to maintain performance under rapid environmental changes.
This leads to a new architectural interpretation of SAGIN as a {hierarchical geometry-aware system}, where:

\begin{itemize}
    \item upper layers ensure availability and large-scale coordination;
    \item lower layers exploit MA to achieve agile local adaptation;
    \item cross-layer interaction is mediated through geometry-aware control.
\end{itemize}

Such a paradigm is particularly critical to LAE services, where system performance depends not only on coverage, but also on the ability to sustain reliable communication, accurate sensing, and safe control under highly dynamic conditions.

\section{MA-Enabled SAGIN for LAE: Layered Vision and Representative Operating Modes}

The discussion so far has clarified that the fundamental limitation of SAGIN in LAE scenarios lies in the mismatch between multi-scale network control and fast local geometry variation. The next step is therefore to understand how MA can be systematically integrated into SAGIN to address this geometry mismatch.

A useful perspective is to view MA-enabled SAGIN as a multi-scale geometry-control hierarchy, in which different layers operate over distinct spatial scopes and timescales, and collectively enable geometry-aware networking. Under this view, the role of MA is inherently non-uniform: rather than being uniformly deployed, antenna mobility should be matched to the dominant geometry impairment and adaptation requirement at each layer.

From a geometry-aware perspective, different layers contribute differently to spatial adaptation. At the macro scale, LEO platforms provide coarse-grained geometry shaping, wide-area coverage, and global coordination. Their role is not to refine individual links, but to maintain large-scale spatial availability. At the meso scale, HAPS platforms provide regional continuity, flexible backhaul support, and intermediate-scale service reinforcement, making them suitable for demand-aware spatial adaptation over quasi-static regions. At the micro scale, UAVs and terrestrial nodes operate in highly dynamic environments, where blockage, interference, and local geometry vary rapidly. It is precisely at this scale that MA becomes most important, since it enables fine-grained, real-time link-level adaptation that cannot be achieved through upper-layer coordination alone.

The resulting design principle can be summarized as scale-aware geometry adaptation: upper layers provide availability, continuity, and large-scale coordination, while lower layers exploit antenna mobility to refine local spatial geometry under fast environmental dynamics. This layered view becomes especially meaningful in LAE scenarios, where multiple functions must be supported simultaneously. Future services will involve not only data transmission, but also safety-critical control, real-time coordination, and sensing-assisted operation. These functions are all highly sensitive to spatial geometry, making geometry-aware adaptation a fundamental requirement rather than an optional enhancement.

\subsection{Operating Mode I: Upper-Layer Support for Aerial Users}

The first representative operating mode arises when the UAV acts primarily as a served aerial node. In this case, the upper layers, namely LEO and HAPS, provide the support required for safe and reliable operation, while the UAV remains the endpoint of that support.

LEO contributes macro-scale coverage and global coordination, which are essential when UAVs traverse large regions or temporarily move beyond local infrastructure. HAPS complements this function by providing regional continuity, lower-latency connectivity, and stable backhaul support. Within this mode, the MA is mounted on the UAV and primarily improves the robustness of the support link received from the upper layers. Rather than replacing platform mobility, UAV-mounted MA refines the local link geometry at a much faster timescale, thereby compensating for body shadowing, attitude-induced beam misalignment, and fast local channel fluctuations. As a result, MA enhances link reliability in scenarios where macro-scale coverage is available but the local reception geometry at the UAV side is unfavorable.
This operating mode is particularly relevant for applications such as long-range logistics, aerial inspection, and mission-critical control, where maintaining continuous and reliable connectivity is more important than maximizing peak throughput. The gain provided by MA is therefore best understood as geometry-driven reliability enhancement.

\subsection{Operating Mode II: Cooperative Service for Ground and Low-Altitude Users}

The second representative operating mode emerges when the UAV acts as a cooperative aerial service node. In this case, LEO and HAPS provide the macro- and meso-scale service structure, together with coordination and backhaul support, while the UAV equipped with MA delivers localized access, relaying, and sensing support to ground users and nearby low-altitude terminals. This mode is particularly important in scenarios such as urban hotspots, infrastructure-poor regions, emergency response, and event-driven service settings.

Here, UAV-mounted MA enables geometry-aware cooperation at the local service side. By adjusting antenna positions, the UAV can adapt its local spatial response to blockage, user distribution, and interference conditions, thereby improving service-side link selection, multi-connectivity, and coordinated transmission. In addition, MA facilitates sensing-assisted service delivery by improving observation geometry, which is critical for applications such as infrastructure monitoring and environmental sensing. Overall, MA makes cooperative service spatially agile, allowing the aerial service node to adapt not only its connectivity, but also its local spatial configuration, to the operating environment.

\subsection{Summary}

Taken together, these operating modes provide a unified interpretation of MA-enabled SAGIN. In the first mode, MA enhances the robustness of aerial users by mitigating local geometry mismatch. In the second, it enables flexible and adaptive cooperation by aligning spatial responses with the environment.
Importantly, these modes are not mutually exclusive. In practical systems, UAVs may dynamically switch between them depending on mission requirements. A UAV may operate as a service object during transit, and later become a cooperative node when serving a local region. This dynamic role transition is particularly characteristic of low-altitude economy scenarios.
From a broader perspective, the layered SAGIN architecture can be interpreted not only as a communication hierarchy, but also as a {geometry-control hierarchy}. In this hierarchy, upper layers provide coarse spatial structure, while MA at the lower layers enables fine-grained, real-time geometry adaptation. This perspective highlights the fundamental role of MAs in transforming spatial geometry from a passive constraint into an actively controllable resource.

\section{Beyond Communication: Multi-Domain Impact of MA-Enabled SAGIN}

This geometry-control capability, as established in the layered SAGIN architecture in the previous section, has profound implications beyond connectivity. 
By enabling fine-grained spatial adaptation at the micro scale while being coordinated with macro- and meso-scale control, MA fundamentally reshapes how different system functions interact with the environment.
The role of MA in SAGIN extends far beyond improving communication metrics. In LAE scenarios, system performance is jointly determined by multiple tightly coupled functionalities, including communication, sensing, control, and navigation. 
A key observation is that all these functions fundamentally depend on the {underlying spatial geometry} of the network, such as the relative positions, orientations, and propagation paths among nodes. However, in conventional systems, this geometry is largely uncontrollable or can only be adjusted at coarse timescales.
MA introduces a new degree of freedom by enabling small-scale antenna displacement or orientation adjustment for real-time geometry adaptation at the link level. 
This capability allows the system not only to optimize communication performance, but also to simultaneously enhance sensing accuracy, control reliability, and navigation integrity. As such, MA provides a mechanism to address multi-domain challenges in SAGIN.

\subsection{Communication: Robustness and Interference Adaptation}

In low-altitude environments, communication links are highly vulnerable to rapid and unpredictable degradation. For instance, UAV motion, building blockage, and platform vibration can cause sudden link quality drops, even when large-scale coverage is ensured.

MA enables the system to actively respond to such micro-scale dynamics. By adjusting antenna positions over a small spatial region, MA can quickly recover from blockage, maintain beam alignment under platform instability, and identify locally favorable propagation paths.
Moreover, MA provides a new mechanism for interference management in 3D environments. Instead of relying solely on beamforming, the system can exploit spatial decorrelation by repositioning antennas, thereby suppressing interference from dynamically varying directions.
Without such fine-grained geometry control, communication performance in SAGIN is often limited by local channel fluctuations, even when sufficient network resources are available.

\subsection{Sensing: Geometry-Aware Perception Enhancement}

Sensing performance in SAGIN is fundamentally constrained by observation geometry. In applications such as SAR imaging, infrastructure monitoring, and target tracking, the relative geometry between the sensing platform and the target directly determines resolution, coverage, and estimation accuracy.
Conventional systems rely on platform movement (e.g., UAV trajectory) to improve sensing geometry. However, such approaches operate at relatively slow timescales and cannot adapt to fast-changing environments or local geometric deficiencies.
MA enable a complementary form of adaptation by adjusting the sensing geometry at a much finer scale. By modifying antenna positions or orientations, MA can enhance angular diversity, improve target observability, and support more flexible bistatic or multistatic sensing configurations.
This capability is particularly valuable in low-altitude scenarios, where obstacles, limited flight paths, and dynamic conditions often restrict the achievable sensing geometry, making fine-grained local adjustment essential.

\subsection{Control: Reliability for Safety-Critical Operations}

Reliable control is a cornerstone of LAE systems, where even short communication outages can lead to severe safety risks. Control links must therefore maintain ultra-high reliability and low latency under highly dynamic conditions.
However, control performance is often limited by local channel degradation, such as blockage or interference, which cannot be effectively mitigated by conventional methods.
MA enhances control reliability by stabilizing the underlying communication links. Through real-time geometry adaptation, MA can reduce the probability of link outages, maintain consistent signal quality, and ensure robust command delivery.
This is particularly critical for applications such as UAV swarm coordination, autonomous navigation, and emergency response, where communication failures can directly compromise system safety.

\subsection{Navigation: Spatial Diversity for Positioning Integrity}

Accurate positioning and navigation are essential for safe and efficient operation in low-altitude environments. However, navigation performance is highly sensitive to the relative geometry between the receiver and its anchors or reference nodes, including their spatial distribution, line-of-sight visibility, and angular diversity. In dense urban or obstructed environments, the available anchors may be concentrated in similar directions or partially blocked, which leads to poor geometric dilution of precision (GDOP), unstable measurements, and reduced positioning integrity even when the received signal strength is still adequate.

MA provides a new approach to improving navigation performance by actively refining the local geometry used for localization. By adjusting antenna positions or orientations over a small local region, the system can improve angular diversity, increase measurement separability, and mitigate adverse propagation conditions such as blockage and multipath distortion. This capability enables more reliable positioning and navigation, which is critical for applications such as urban air mobility, autonomous delivery, and airspace management in the LAE.

\subsection{Summary}

The above discussions reveal a fundamental insight: communication, sensing, control, and navigation are not independent functions, but are all governed by the underlying spatial geometry of the system.
Conventional SAGIN designs treat these functions separately and rely on coarse-grained adaptation mechanisms, leading to suboptimal performance in highly dynamic environments.

MAs enable a paradigm shift toward {multi-function geometry-aware networking}, where the system can jointly optimize its spatial configuration to balance multiple objectives. This unified perspective allows SAGIN to achieve both large-scale coverage and fine-grained adaptability, which are essential for supporting diverse LAE services.
In this sense, MA transforms spatial geometry from a passive constraint into an actively controllable resource, fundamentally redefining how wireless systems are designed and operated.

\section{Illustrative Numerical Insights for MA-Enabled SAGIN}

To provide intuitive validation of the proposed geometry-aware SAGIN framework, we present two representative numerical examples that highlight the impact of MA on system robustness and multi-functional performance. The goal of these results is not to claim optimality, but to illustrate how geometry control fundamentally improves system behavior under practical low-altitude conditions.

\subsection{Multi-Scale Coverage Robustness}

We first examine the robustness of coverage under increasing blockage conditions, which is one of the dominant impairments in low-altitude environments. The blockage probability is varied to emulate increasingly challenging propagation conditions, and the outage probability is used as the performance metric.

\begin{figure}
\centering
\includegraphics[width=0.48\textwidth]{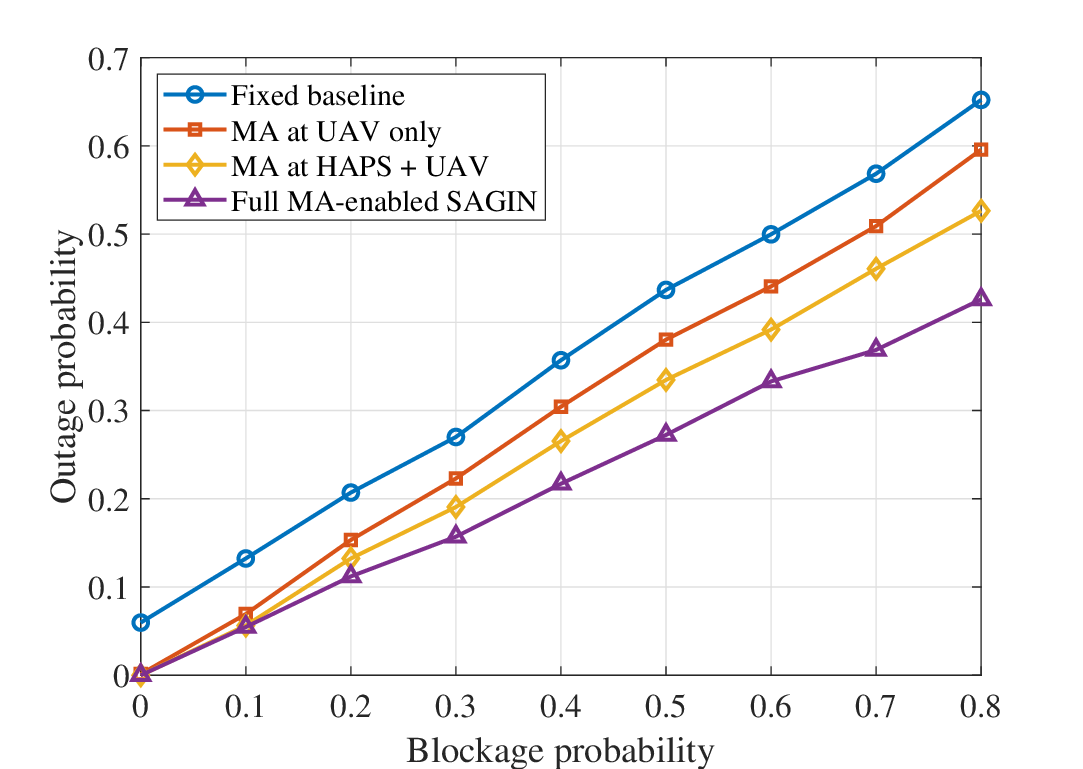}
\caption{Multi-scale coverage robustness of MA-enabled SAGIN under increasing blockage probability. The outage probability is evaluated for different deployment configurations, including a fixed baseline, MA at the UAV layer only, MA at both HAPS and UAV layers, and a full MA-enabled SAGIN. 
}
\label{fig:coverage}
\end{figure}

Fig.~\ref{fig:coverage} compares four configurations: a fixed baseline without antenna mobility, MA deployed at the UAV layer only, MA deployed at both HAPS and UAV layers, and a fully MA-enabled SAGIN architecture.

It can be observed that, as the blockage probability increases, all schemes experience performance degradation. However, the degradation behavior differs significantly across configurations. The fixed baseline suffers from a rapid increase in outage probability, reflecting its inability to cope with unfavorable geometry conditions. Even without blockage, the fixed baseline still suffers from unfavorable static geometry and lacks local spatial adaptation capability. In contrast, MA-enabled schemes exhibit a much more gradual degradation.

The observed performance gain is closely related to the role of MA in enabling fine-grained spatial adaptation at the micro scale. Instead of relying solely on large-scale coverage, the system can dynamically adjust its local geometry to maintain favorable link conditions. In addition, the progressively improved performance from UAV-only MA to multi-layer MA deployment highlights the importance of coordinated geometry control across different layers.

\subsection{Communication--Sensing Tradeoff in Safety-Critical Operation}

We next investigate the tradeoff between communication performance and sensing accuracy, which is a key requirement in LAE services involving integrated sensing and communication.

\begin{figure}
\centering
\includegraphics[width=0.4\textwidth]{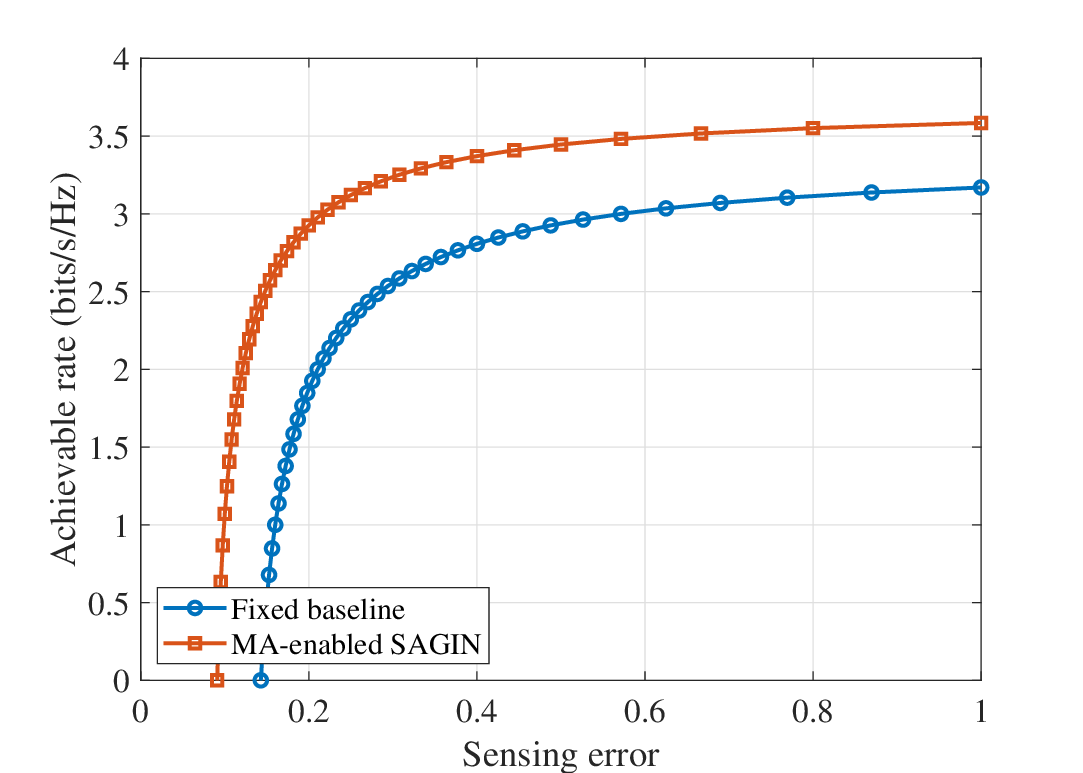}
\caption{Communication--sensing tradeoff for fixed and MA-enabled SAGIN architectures. The achievable rate is plotted against sensing error by varying the system emphasis between communication and sensing objectives. 
}
\label{fig:tradeoff}
\end{figure}

Fig.~\ref{fig:tradeoff} shows the achievable rate versus sensing error for both the fixed baseline and the MA-enabled SAGIN. The tradeoff is generated by varying the system emphasis between communication and sensing objectives.
It can be seen that the MA-enabled scheme consistently outperforms the fixed baseline across the entire tradeoff region. In particular, for the same sensing accuracy, MA achieves a higher communication rate, while for the same communication performance, it achieves lower sensing error.

This result indicates that MA enables a more efficient balance between communication and sensing objectives. By providing additional spatial degrees of freedom, the system can better align its geometry with both signal transmission and target observation requirements. As a result, the achievable performance region is expanded, particularly in regimes where both reliability and perception accuracy are critical.

\subsection{Key Takeaways}

The above results reveal several important insights regarding MA-enabled SAGIN:

\begin{itemize}
    \item \textbf{Enhanced robustness under geometry degradation}: MA significantly improves coverage reliability in blockage-dominated environments by enabling local geometry adaptation.
    
    \item \textbf{Improved multi-function tradeoffs}: MA enlarges the achievable communication--sensing tradeoff region, highlighting its role in supporting integrated multi-domain services.
    
    \item \textbf{Effectiveness of multi-scale deployment}: The performance gain becomes more pronounced when MA is deployed across multiple layers, validating the proposed geometry-control hierarchy in SAGIN.
\end{itemize}

\section{Current Design Challenges, Emerging Solutions, and Future Directions}

The preceding sections have shown that MA-enabled SAGIN can significantly improve robustness and multi-functional performance by enabling geometry-aware adaptation across layers. The illustrative numerical results further support this potential under representative low-altitude settings. However, these gains have so far been demonstrated under simplified assumptions. Translating this vision into a practical and scalable system requires addressing a number of fundamental challenges, including how geometry control can be implemented, coordinated, and validated across heterogeneous platforms with different mobility patterns, service scopes, and operational constraints.

\subsection{Current Design Challenges and Emerging Solutions}

\textbf{Platform-dependent MA design.}
MA cannot be implemented in a platform-agnostic manner, since LEO, HAPS, UAV, and terrestrial nodes operate under fundamentally different constraints in reliability, payload, power, endurance, and maintenance. While LEO platforms emphasize large-scale stability and coverage shaping, HAPS naturally support regional persistence and elastic service adaptation. In contrast, UAVs must realize antenna mobility under stringent limitations in payload and energy, while still responding to rapid geometry variation. At the terrestrial layer, the focus shifts to cost-efficient integration with existing infrastructure and multiuser processing. 

A practical solution is to adopt a {platform-specific co-design philosophy}, where movable range, actuation speed, and control granularity are matched to the geometry-control role of each layer, rather than enforcing a uniform hardware abstraction.

\medskip

\textbf{Multi-timescale control and network orchestration.}
Among all challenges, a particularly critical one is how to coordinate geometry adaptation across layers operating on vastly different timescales. LEO and HAPS naturally handle slow, large-scale spatial adaptation driven by orbital or quasi-stationary dynamics, whereas UAV and terrestrial layers must react to fast local variations such as blockage, interference, and user mobility.

This mismatch suggests that MA-enabled SAGIN should adopt a {hierarchical orchestration structure}, where coarse geometry shaping is handled at upper layers, and fine-grained, real-time geometry refinement is pushed toward lower layers. A practical direction is to combine event-triggered adaptation, lightweight local control, and limited cross-layer feedback, instead of relying on fully centralized coordination.

\medskip

\textbf{Protocol hooks and cross-layer signaling.}
To translate antenna-level flexibility into system-level gains, MA must be visible beyond the physical layer. This requires lightweight descriptors of movable range, actuation capability, and geometry-related indicators such as blockage and interference conditions. 

Rather than exposing detailed antenna control, a practical approach is to abstract MA into a small set of network-facing states, enabling its integration into link selection, multi-connectivity, and cooperative transmission decisions. In this way, geometry control becomes part of cross-layer decision loops rather than an isolated function.

\medskip

\textbf{Validation and evaluation frameworks.}
The benefits of MA-enabled SAGIN cannot be fully captured by link-level analysis alone, as the architecture inherently couples antenna mobility with geometry evolution, control signaling, and multi-layer coordination. Realistic evaluation therefore requires system-level emulation, motion-aware channel modeling, and hardware-in-the-loop experimentation.

A promising pathway is to develop hierarchical validation frameworks that combine analytical abstraction, system-level simulation, and limited-scale field testing, allowing geometry-aware behavior to be assessed under realistic conditions.

\subsection{Future Challenges and Possible Solutions}

Looking ahead, several additional challenges will shape the evolution of MA-enabled SAGIN. These challenges, while diverse in form, share a common origin: the need to sense, predict, and control spatial geometry across multiple layers and timescales.

\medskip

\textbf{Scalable channel awareness.}
Once antenna position becomes part of the control loop, the channel becomes inherently position-dependent, leading to strong coupling between channel acquisition, spatial adaptation, and system control. 
A promising direction is to combine geometry-aware channel prediction and lightweight tracking with learning-assisted abstraction. In particular, data-driven models can capture complex position-dependent channel variations and support fast spatial decision making, enabling efficient spatial awareness without excessive overhead.

\medskip

\textbf{Backhaul-aware multi-layer coordination.}
As MA improves local adaptability, system bottlenecks may shift toward upper-layer coordination and backhaul constraints. In many scenarios, access quality alone is insufficient to ensure service continuity. This motivates joint access--backhaul design, where antenna mobility and multi-layer resource allocation are optimized together.

\medskip

\textbf{Communication--sensing convergence.}
Future low-altitude systems will increasingly rely on joint awareness of the environment. In this context, MA becomes a tool not only for link enhancement, but also for sensing geometry control. This raises a new challenge: how to coordinate antenna mobility when communication and sensing objectives are partially aligned but not identical. A promising solution is task-aware MA control, where system objectives are defined at the mission level rather than per-function metrics.

\medskip

\textbf{Standardization and benchmarking.}
The lack of common evaluation frameworks currently limits fair comparison and practical adoption. Establishing benchmark scenarios for low-altitude corridors, urban blockage environments, and emergency services will be essential. Such benchmarks should capture both communication and sensing objectives, enabling systematic evaluation of geometry-aware designs.

\medskip

\textbf{Energy-aware MA control.}
While MA introduces a valuable spatial degree of freedom, it also incurs non-negligible actuation and control overhead, particularly for resource-constrained platforms such as UAVs and HAPS. In low-altitude scenarios, antenna mobility must be carefully balanced against payload limitations, onboard energy budgets, and mission duration, creating a fundamental tradeoff between geometry adaptation performance and energy efficiency. A promising direction is to develop {energy-aware MA control strategies}, where antenna movement is selectively triggered based on environmental dynamics, service urgency, and expected performance gain. By combining lightweight decision rules, event-driven adaptation, and predictive modeling, the system can achieve efficient geometry control while avoiding unnecessary actuation, enabling practical deployment without compromising energy sustainability.

\subsection{Discussion}

Taken together, these challenges highlight that the evolution of MA-enabled SAGIN is not merely a matter of enhancing antenna capabilities. Rather, it requires a comprehensive system-level perspective that integrates platform-aware design, hierarchical control, protocol adaptation, and realistic validation.
More fundamentally, these issues reflect a broader shift: spatial geometry is no longer a passive constraint, but an actively controllable resource. Realizing this vision will be key to transforming MA-enabled SAGIN from a conceptual framework into practical infrastructure for LAE services.

\section{Conclusion}

MAs offer a fundamentally new perspective for SAGIN. By introducing antenna mobility as an additional spatial degree of freedom, MA goes beyond conventional beamforming and enables direct interaction with the underlying propagation geometry. Rather than merely improving individual links, MA provides a mechanism for reshaping how heterogeneous layers coordinate, adapt, and respond to dynamic environments.
In this article, we have argued that MA-enabled SAGIN is particularly well suited for LAE services, where wide-area availability, regional continuity, and local robustness must be jointly supported. We presented a layered architecture, discussed representative operating modes, provided illustrative numerical insights, and outlined key challenges toward practical deployment.
More importantly, MA-enabled SAGIN points toward a broader paradigm shift: spatial geometry is no longer a passive constraint, but an actively controllable resource. From this perspective, MA should be understood not simply as an antenna enhancement, but as a fundamental enabler of geometry-aware, multi-scale, and multi-functional network design for future non-terrestrial systems.


\begin{thebibliography}{10}
	\providecommand{\url}[1]{#1}
	\csname url@samestyle\endcsname
	\providecommand{\newblock}{\relax}
	\providecommand{\bibinfo}[2]{#2}
	\providecommand{\BIBentrySTDinterwordspacing}{\spaceskip=0pt\relax}
	\providecommand{\BIBentryALTinterwordstretchfactor}{4}
	\providecommand{\BIBentryALTinterwordspacing}{\spaceskip=\fontdimen2\font plus
		\BIBentryALTinterwordstretchfactor\fontdimen3\font minus
		\fontdimen4\font\relax}
	\providecommand{\BIBforeignlanguage}[2]{{%
			\expandafter\ifx\csname l@#1\endcsname\relax
			\typeout{** WARNING: IEEEtran.bst: No hyphenation pattern has been}%
			\typeout{** loaded for the language `#1'. Using the pattern for}%
			\typeout{** the default language instead.}%
			\else
			\language=\csname l@#1\endcsname
			\fi
			#2}}
	\providecommand{\BIBdecl}{\relax}
	\BIBdecl
	
	\bibitem{zeng2019accessing}
	Y.~Zeng, Q.~Wu, and R.~Zhang, ``Accessing from the sky: A tutorial on uav
	communications for 5g and beyond,'' \emph{Proceedings of the IEEE}, vol. 107,
	no.~12, pp. 2327--2375, 2019.
	
	\bibitem{mahboob2024revolutionizing}
	S.~Mahboob and L.~Liu, ``Revolutionizing future connectivity: A contemporary
	survey on ai-empowered satellite-based non-terrestrial networks in 6g,''
	\emph{IEEE Communications Surveys \& Tutorials}, vol.~26, no.~2, pp.
	1279--1321, 2024.
	
	\bibitem{liu2018space}
	J.~Liu, Y.~Shi, Z.~M. Fadlullah, and N.~Kato, ``Space-air-ground integrated
	network: A survey,'' \emph{IEEE Communications Surveys \& Tutorials},
	vol.~20, no.~4, pp. 2714--2741, 2018.
	
	\bibitem{xu2023space}
	J.~Xu, M.~A. Kishk, and M.-S. Alouini, ``{Space-air-ground-sea integrated
		networks: Modeling and coverage analysis},'' \emph{IEEE Transactions on
		Wireless Communications}, vol.~22, no.~9, pp. 6298--6313, 2023.
	
	\bibitem{qin2023coverage}
	Y.~Qin, M.~A. Kishk, and M.-S. Alouini, ``{Coverage analysis and trajectory
		optimization for aerial users with dedicated cellular infrastructure},''
	\emph{IEEE Transactions on Wireless Communications}, vol.~23, no.~4, pp.
	3042--3056, 2023.
	
	\bibitem{wu2019towards}
	Q.~Wu and R.~Zhang, ``{Towards smart and reconfigurable environment:
		Intelligent reflecting surface aided wireless network},'' \emph{IEEE
		communications magazine}, vol.~58, no.~1, pp. 106--112, 2019.
	
	\bibitem{di2020smart}
	M.~Di~Renzo, A.~Zappone, M.~Debbah, M.-S. Alouini, C.~Yuen, J.~De~Rosny, and
	S.~Tretyakov, ``{Smart radio environments empowered by reconfigurable
		intelligent surfaces: How it works, state of research, and the road ahead},''
	\emph{IEEE journal on selected areas in communications}, vol.~38, no.~11, pp.
	2450--2525, 2020.
	
	\bibitem{zhu2023modeling}
	L.~Zhu, W.~Ma, and R.~Zhang, ``Modeling and performance analysis for movable
	antenna enabled wireless communications,'' \emph{IEEE Transactions on
		Wireless Communications}, vol.~23, no.~6, pp. 6234--6250, 2023.
	
	\bibitem{zhu2025tutorial}
	L.~Zhu, W.~Ma, W.~Mei, Y.~Zeng, Q.~Wu, B.~Ning, Z.~Xiao, X.~Shao, J.~Zhang, and
	R.~Zhang, ``A tutorial on movable antennas for wireless networks,''
	\emph{IEEE Communications Surveys \& Tutorials}, 2025.
	
	\bibitem{shao2025tutorial}
	X.~Shao, W.~Mei, C.~You, Q.~Wu, B.~Zheng, C.-X. Wang, J.~Li, R.~Zhang,
	R.~Schober, L.~Zhu \emph{et~al.}, ``A tutorial on six-dimensional movable
	antenna for 6g networks: Synergizing positionable and rotatable antennas,''
	\emph{IEEE Communications Surveys \& Tutorials}, 2025.
	
\end{thebibliography}

\end{document}